\documentclass{ws-ijmpe}
\usepackage{tensor}
\usepackage{psfrag}
\usepackage{bm}

\renewcommand{\vec}[1]{\bm{#1}}
\def\Im{\mathop{\rm Im}\nolimits}

\begin{document}
\markboth{D. Fern\'andez-Fraile \& A. G\'omez Nicola}{Transport properties of a meson gas}

\catchline{}{}{}{}{}

\title{TRANSPORT PROPERTIES OF A MESON GAS}

\author{D. FERN\'ANDEZ-FRAILE\footnote{\texttt{danfer@fis.ucm.es}}  ~and A. G\'OMEZ NICOLA\footnote{\texttt{gomez@fis.ucm.es}}}

\address{Departamentos de F\'isica Te\'orica I y II, Universidad Complutense, 28040 Madrid, Spain}

\maketitle

\begin{history}
\received{(24 May 2007)}
\accepted{(12 June 2007)}
\end{history}

\begin{abstract}
We present recent results on a systematic method to calculate
transport coefficients for a meson gas (in particular, we analyze a
pion gas) at low temperatures in the context of Chiral Perturbation
Theory. Our method is based on the study of Feynman diagrams with a
power counting which takes into account collisions in the plasma by
means of a non-zero particle width. In this way, we obtain results
compatible with analysis of Kinetic Theory  with just the leading
order diagram. We show the behavior with temperature of electrical
and thermal conductivities and shear and bulk viscosities, and we
discuss the fundamental role played by unitarity. We obtain that
bulk viscosity is negligible against shear viscosity near the chiral
phase transition. Relations between the different transport
coefficients and bounds on them based on different theoretical
approximations are also discussed. We also comment on some
applications to heavy-ion collisions.
\end{abstract}

\section{Introduction}

Chiral Perturbation Theory (ChPT) is an effective field theory which
describes the low-energy meson dynamics,  based on the spontaneous
breaking of the QCD chiral symmetry. The ChPT lagrangian consists of
an expansion in masses and derivatives of the meson fields and
perturbation theory is performed in powers of $p$ (a meson momentum,
mass or  temperature) over $\mathnormal{\Lambda}_\chi\sim 1\
\mathrm{GeV}$ (for masses and momenta) or the critical temperature
$T_\mathrm{c}\sim 200\ \mathrm{MeV}$ (for temperatures). At zero
temperature, the order $\mathcal{O}(p^D)$ of a  Feynman diagram is
estimated  by Weinberg's power counting: $D = 2+\sum_n N_n (n-2) + 2
L$, where $L$ is the number of loops, and $N_n$ is the number of
vertices coming  from the lagrangian with $n$ derivatives. We will
see that, for the calculation of transport coefficients, this
formula does not provide the correct order for the contribution of
the dominant diagrams.

In Linear Response Theory (LRT), a transport coefficient
$\mathfrak{T}$ is calculated by means of Kubo-like
expressions:\cite{FN}
\begin{equation}\label{transport}
\mathfrak{T}=C_\mathfrak{T}\lim_{q^0\rightarrow 0^+}\lim_{|\vec{q}|\rightarrow 0^+}\frac{\partial}{\partial{q^0}}\Im \mathrm{i}\int\mathrm{d}^4x\ \mathrm{e}^{\mathrm{i}q\cdot x}\theta(t)\langle[\mathcal{A}_\mathfrak{T}(x),\mathcal{A}_\mathfrak{T}(0)]\rangle\ ,
\end{equation}
where $C_\mathfrak{T}$ is a constant, and $\mathcal{A}_\mathfrak{T}$
certain operator (tensorial in general). The calculation of
transport coefficients is non-perturbative due to the presence in
the diagrams of pair of lines with \emph{non-zero width} sharing the
same four-momentum. The width corresponds to the inverse of the mean
collision time. These lines give rise to products of the kind
$G_\mathrm{A} G_\mathrm{R}\propto 1/\mathnormal{\Gamma}$, with
$\mathnormal{\Gamma}$ the particle width. According to this, in
principle, the  dominant contribution to transport coefficients
would come from  the so called \emph{ladder and bubble diagrams}. In
order to take into account these non-perturbative contributions, it
can be shown\cite{FN} that in ChPT is necessary to use a power
counting different from Weinberg's one for diagrams with
$\mathnormal{\Gamma}\neq 0$. The leading contribution $\mathfrak{T}^{(0)}$ at low
temperatures corresponds to a single bubble diagram with nonzero
width. For $T\gtrsim M_\pi$ it may be necessary to sum the ladder
diagrams due to the dominance of derivative vertices.\cite{FN}

In the dilute gas approximation for the pion gas,
$\mathnormal{\Gamma}$ depends linearly on $\sigma_{\pi\pi}$, the
pion-pion scattering cross section, that can be written in terms of
partial waves $t_{IJ}$ of total isospin $I$ and angular momentum
$J$. A crucial point is that the behavior of transport coefficients
with increasing temperature is directly linked with that of the
partial waves with increasing energy and therefore the unitarity
bound plays a fundamental role.  ChPT violates the unitarity
condition $\mathrm{S}\mathrm{S}^\dagger=\hat{1}\Rightarrow \Im
t_{IJ}(s)=\sigma(s)\left|t_{IJ}(s)\right|^2$, with
$\sigma(s)\equiv\sqrt{1-4M_\pi^2/s}$ the $\pi\pi$ phase space, since
partial waves contain polynomials in $s$.
We will use\cite{FN} the  so called Inverse Amplitude Method, which
provides a exactly unitary amplitude reproducing the chiral
expansion near the two-pion threshold.

\section{Electrical and thermal conductivities}

\begin{figure}
\begin{center}
\parbox[c]{6.2cm}{
\psfrag{ylabel}[][][0.8]{$\displaystyle\frac{\sigma^{(0)}}{M_\pi}$}
\psfrag{xlabel}[][][0.8]{$T\ (\mathrm{MeV})$}
\psfrag{unit}[l][l][0.6]{unitarized $\mathcal{O}(p^4)$}
\psfrag{Op4}[l][l][0.6]{$\mathcal{O}(p^4)$}
\psfrag{lowtemp}[l][l][0.7]{$\displaystyle\sim\frac{1}{\sqrt{T}}$}
\includegraphics[width=6.2cm]{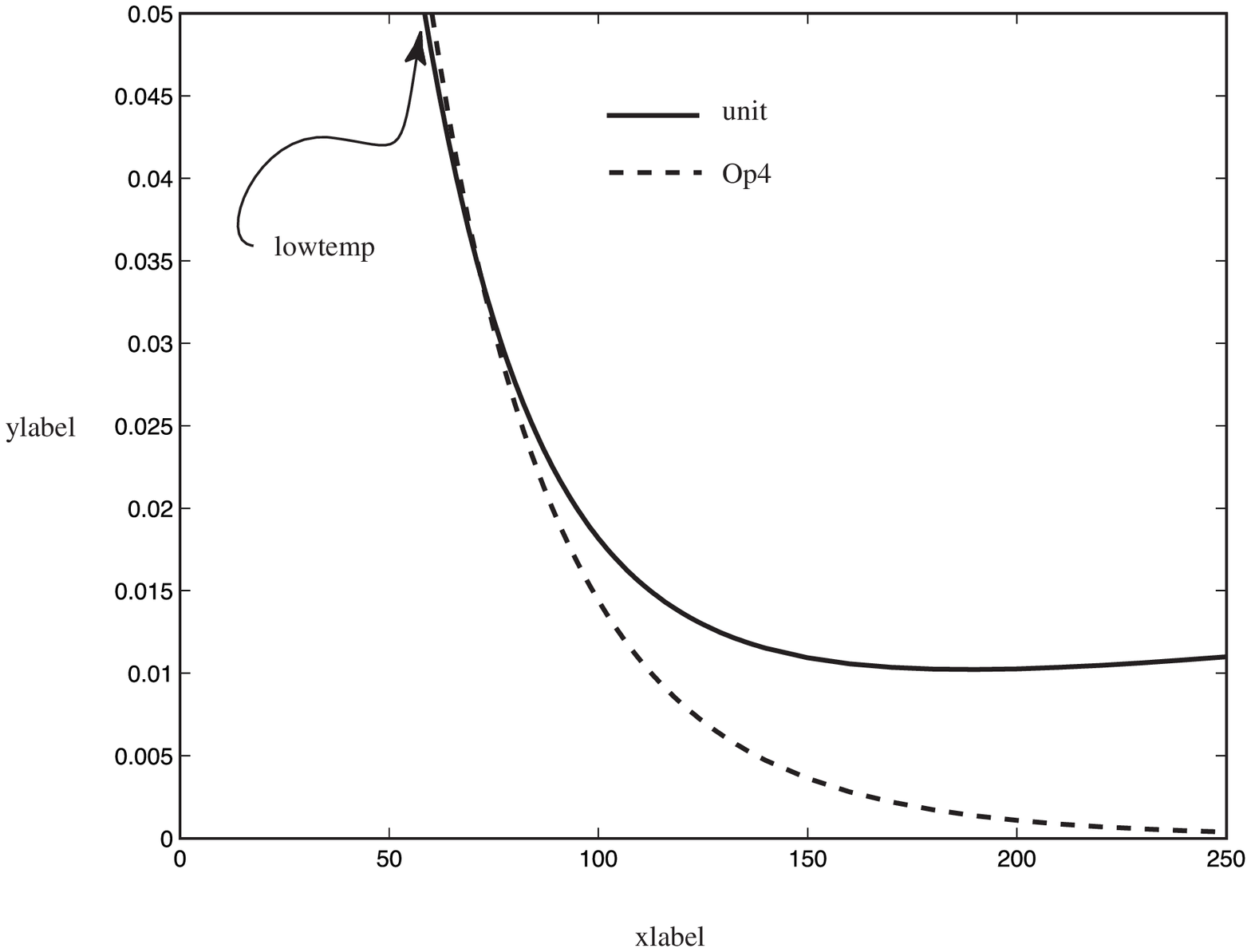}}
\hspace{0.1cm}\parbox[c]{6.2cm}{
\psfrag{lowtemp}[l][l][0.7]{$\displaystyle\sim\frac{1}{T^{3/2}}$}
\psfrag{unit}[l][l][0.6]{unitarized $\mathcal{O}(p^4)$}
\psfrag{Op4}[l][l][0.6]{$\mathcal{O}(p^4)$}
\psfrag{Torres}[l][l][0.6]{Torres \textit{et
al.},\cite{Torres} IAM}
\psfrag{Prakash}[l][l][0.6]{Prakash \textit{et
al.},\cite{Prakash} 3rd order}
\psfrag{xlabel}[][][0.8]{$T\ (\mathrm{MeV})$}
\psfrag{ylabel}[][][0.8]{\parbox{1cm}{\begin{center}$\displaystyle\kappa^{(0)}$\\$\displaystyle(\mathrm{GeV}^2)$\end{center}}}
\includegraphics[width=6.2cm]{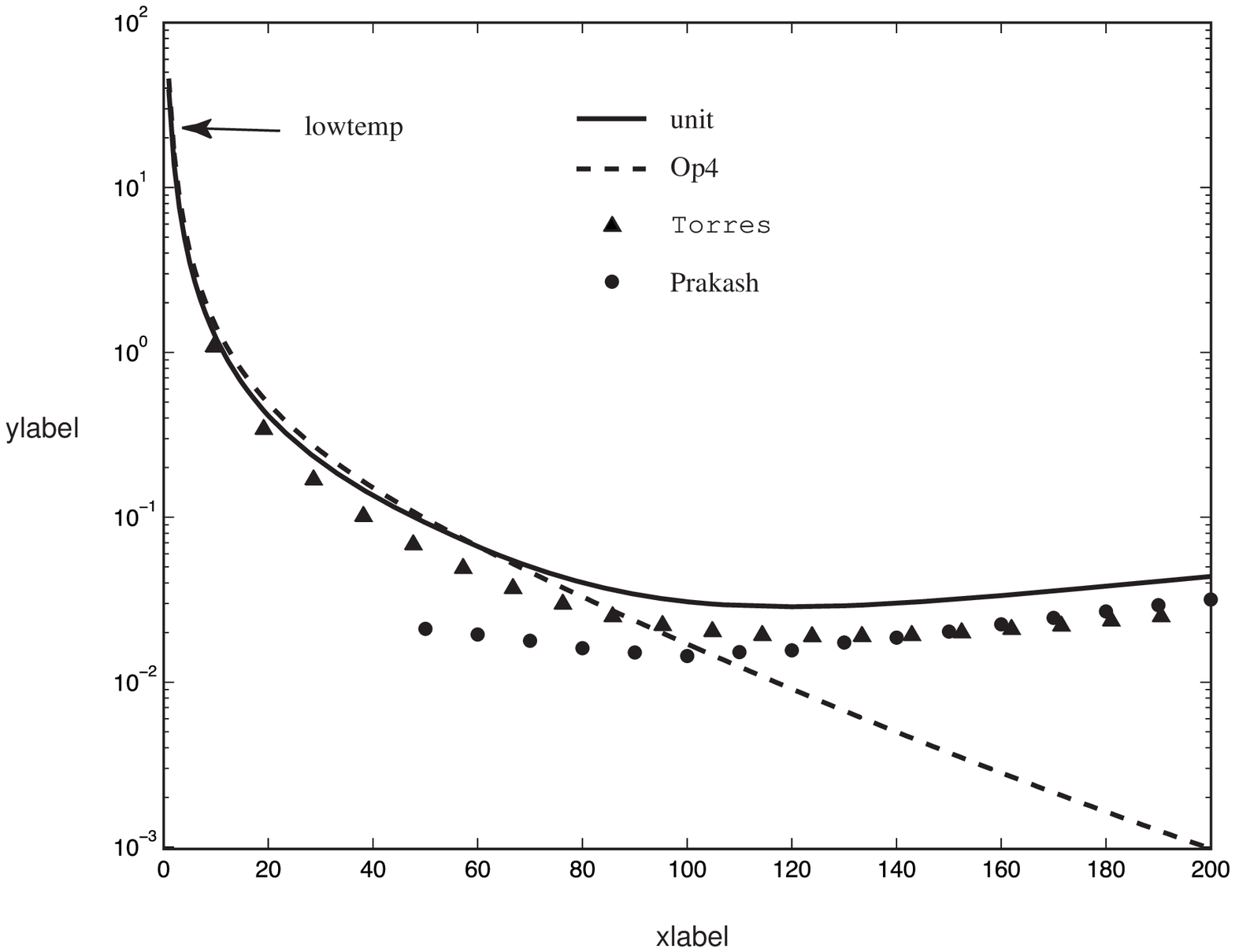}}
\end{center}
\caption{(Left) Results for the DC conductivity. (Right) Results for the thermal conductivity.}\label{figcond}
\end{figure}

Specifically, we analyze the response from a gas against the
presence of a constant electric field (DC conductivity). In this
case in (\ref{transport}) we have $C_\sigma\equiv-1/3$ and
$\mathcal{A}_\sigma\equiv J^i$ (electric current). According to
Kinetic Theory (KT) $\displaystyle\sigma\sim
e^2n_\mathrm{ch}\tau/M_\pi$ ($n_\mathrm{ch}$ is the density of
charged particles, $\tau$ is the collision mean time, and $e$ is the
particle charge), and $\tau\sim1/\mathnormal{\Gamma}$,
$\mathnormal{\Gamma}\sim n v\sigma_{\pi\pi}$ ($v$ is the mean speed
of the particles). In the nonrelativistic limit, $n\sim (\sqrt{M_\pi
T})^3\mathrm{e}^{-M_\pi/T}$, $v\sim\sqrt{T/M_\pi}$, and
$\sigma_{\pi\pi}$ is a constant, so that $\sigma\sim 1/\sqrt{T}$.
This is consistent with our result in the  ChPT diagrammatic
approach for $T\ll M_\pi$, namely $\sigma^{(0)}\simeq 15\ e^2
F_\pi^4 T^{-1/2} M_\pi^{-5/2}$ ($F_\pi$ is the pion decay constant).
The results for higher $T$ are shown in Fig. \ref{figcond} (left)
where we see that unitarity changes the decreasing behaviour of the
DC conductivity.

The electrical conductivity is related to the soft-photon spectrum
emitted by the gas of pions. Considering an hydrodynamical model of
spherical symmetry in order to describe the expansion of the gas
after a relativistic heavy ion collision,\cite{FN} we obtain a
prediction at the origin for the photon spectrum compatible with
experimental results.\cite{FN,WA98}

For the thermal conductivity, $C_\kappa\equiv -1/(3T)$, and
$\mathcal{A}_\kappa\equiv T_{0i}$. By KT: $\kappa\sim c_p l v$
($c_p$ is the heat capacity per unit volume and $\displaystyle l\sim
1/(\sigma_{\pi\pi} n)$ is the particle mean free path). In the
nonrelativistic limit, $c_p\sim T^{-1/2}\mathrm{e}^{-M_\pi/T}$ so
that $\kappa\sim T^{-3/2}$. In our ChPT scheme we obtain: for $T\ll
M_\pi$, $\kappa^{(0)}\simeq 10\ F_\pi^4 T^{-3/2}M_\pi^{-1/2}$, and
for $T\simeq M_\pi$, $T\cdot\kappa^{(0)}\sim\eta^{(0)}$. In Fig.
\ref{figcond} (right) we observe that our results for $\kappa$ get
very close to those obtained by a KT analysis\cite{Torres} with
unitarized amplitudes from ChPT.

\section{Shear and bulk viscosities}

\begin{figure}
\begin{center}
\parbox[c]{6.2cm}{
\psfrag{ylabel}[c][c][0.8]{$(\mathrm{GeV}^3)$}
\psfrag{xlabel}[][][0.8]{$T\ (\mathrm{MeV})$}
\psfrag{shearunit}[c][c][0.5]{$\eta^{(0)}$, unitarized $\mathcal{O}(p^4)\quad{}$}
\psfrag{shearPrakash}[c][c][0.5]{$\eta$, Prakash 3rd order}
\psfrag{shearOp4}[c][c][0.5]{$\eta^{(0)}$, $\mathcal{O}(p^4)$}
\psfrag{bulkunit}[c][c][0.5]{$\zeta^{(0)}$, unitarized $\mathcal{O}(p^4)\quad{}$}
\psfrag{bulkPrakash}[c][c][0.5]{$\zeta$, Prakash 3rd order}
\psfrag{bulkOp4}[c][c][0.5]{$\zeta^{(0)}$, $\mathcal{O}(p^4)$}
\psfrag{lowtemp}[c][c][0.6]{$\sim\sqrt{T}$}
\includegraphics[width=6.2cm]{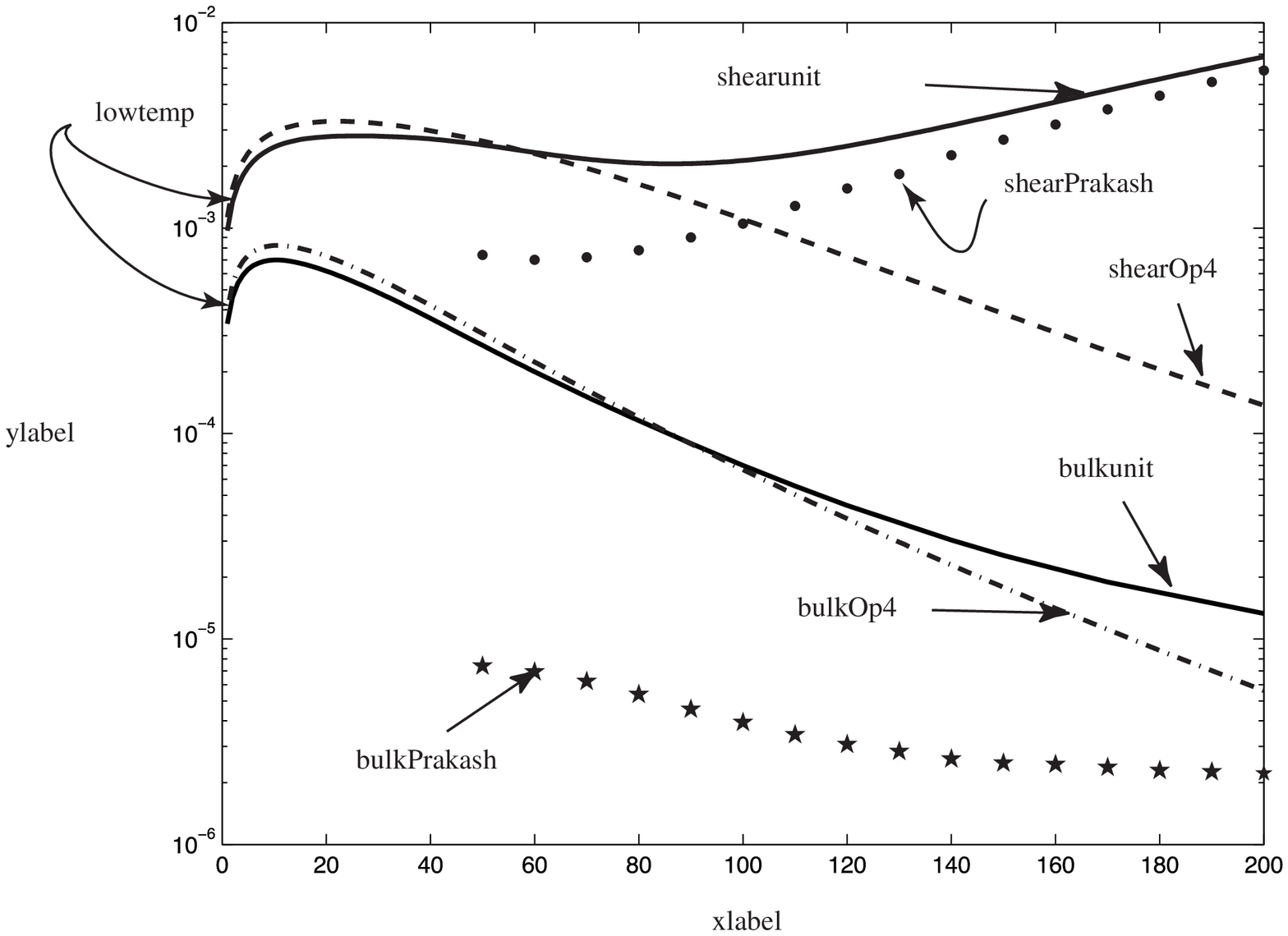}}
\hspace{0.1cm}\parbox[c]{6.2cm}{
\psfrag{ylabel}[cc][cc][0.8]{$\displaystyle\frac{\eta^{(0)}}{s}$}
\psfrag{xlabel}[][][0.8]{$T\ (\mathrm{MeV})$}
\psfrag{unit}[l][l][0.6]{unitarized $\mathcal{O}(p^4)$}
\psfrag{Op4}[l][l][0.6]{$\mathcal{O}(p^4)$}
\psfrag{bound}[l][l][0.6]{$1/(4\pi)$}
\includegraphics[width=6.2cm]{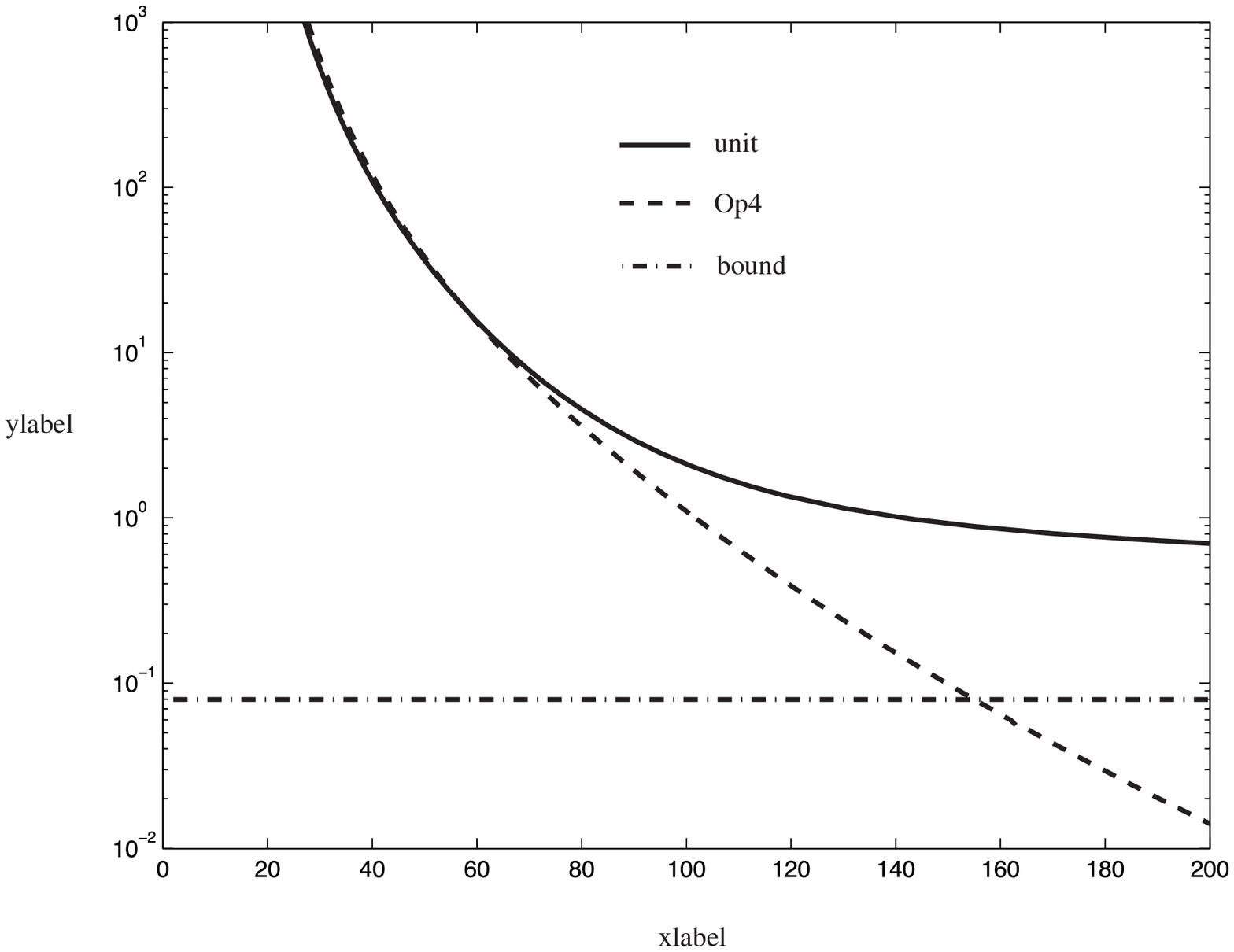}}
\end{center}
\caption{(Left) Shear and bulk viscosities against temperature. (Right) $\eta/s$ quotient.}\label{figvisco}
\end{figure}

The shear viscosity, $\eta$, is calculated by means of the formula
(\ref{transport}) with $C_\eta\equiv 1/10$ and
$\mathcal{A}_\eta\equiv T_{ij}-g_{ij}\tensor{T}{^l_l}/3$
(non-diagonal part of the energy-momentum tensor). For the bulk
viscosity, $\zeta$, $C_\zeta\equiv 1$, and
$\mathcal{A}_\zeta\equiv-\tensor{T}{^l_l}/3-v_\mathrm{s}^2 T_{00}$,
where $v_\mathrm{s}$ is the speed of sound in the gas. In Fig.
\ref{figvisco} (left)  we compare our results for viscosities with
those obtained by Prakash \textit{et al.} using a KT
analysis.\cite{Prakash} We also agree with a pion gas KT work by
Dobado \textit{et al.}\cite{Dobado} By nonrelativistic KT we expect
now the behavior $\eta,\zeta\sim M_\pi v n l$, so that
 $\eta,\zeta\sim\sqrt{T}$ and both viscosities should then be of the same order
at very low temperatures. In fact, our ChPT calculation gives for
$T\ll M_\pi$, $\eta^{(0)}\simeq 37\ T^{1/2}F_\pi^4 M_\pi^{-3/2}$,
and $\zeta^{(0)}\simeq 0.36\ \eta^{(0)}$ (also compatible with the
analysis of Gavin\cite{Gavin}). For $T\simeq M_\pi$, $\eta^{(0)}\sim
1000\ \zeta^{(0)}$, so that the bulk viscosity is negligible at
temperatures accessible in experiments. In fact, we get
$\zeta^{(0)}\sim (1/3-v_\mathrm{s}^2)^2\eta^{(0)}\rightarrow 0$ for
$T\gg M_\pi$, in agreement with Hosoya \textit{et al.}\cite{Hosoya}
and Arnold \textit{et al.}\cite{Arnold}


Unitarity makes the quotient $\eta/s$ ($s$ is the entropy density)
for the pion gas respect the bound $1/(4\pi)$ predicted by Kovtun
\textit{et al.},\cite{Kovtun} as we can see in Fig. \ref{figvisco}
(right). In addition, near $T_c$ our value for
$\eta/s$ is not far from  recent lattice and model
estimates.\cite{Nakamura}   Although we do not represent it in the
Figure, we do obtain a behavior for $\eta/s$ growing with $T$ for
temperatures (unrealistic) $>550\ \mathrm{MeV}$. A slowly increasing
behavior is also obtained by calculations from the quark gluon
plasma (QGP) phase.\cite{Kapusta} The sound attenuation length is
given by (neglecting the contribution from the bulk viscosity)
$\mathnormal{\Gamma}_\mathrm{s}\simeq 4\eta/(3sT)$, and is directly
related to phenomenological effects such as \emph{elliptic flow} or
\emph{HBT radii}. We get, at $T=180\ \mathrm{MeV}$, the value
$\mathnormal{\Gamma}_\mathrm{s}\simeq 1.1\ \mathrm{fm}$, in
agreement with the estimate of Teaney.\cite{Teaney}

\section*{Acknowledgements}
We would like to thank A. Dobado, F. J. Llanes-Estrada, and J. M. Torres Rinc\'on for useful comments. We also thank the financial support from the Spanish research projects FPA2004-02602, PR27/05-13955-BSCH, FPA2005-02327 and the doctoral fellowship BES-2005-6726 from the MEC-FPI programme.

\end{document}